\documentclass[aps,prc,reprint,nofootinbib]{revtex4-1}

\usepackage{graphicx}   %       for graphics
\usepackage{latexsym}   %       for special symbols
\usepackage{enumerate}
\usepackage{tabularx}

\begin{document}

\title{From Cosmic Matter to the Laboratory}

\author{Anton Motornenko$^{1,2}$, Jan Steinheimer$^{2}$, Horst Stoecker$^{1,2,3}$}

\affiliation{$^{1}$Institut für Theoretische Physik,
Johann Wolfgang Goethe-Universität Frankfurt am Main, Germany}
\affiliation{$^{2}$Frankfurt Institute for Advanced Studies, Giersch Science Center, Frankfurt am Main, Germany}
\affiliation{$^{3}$Gesellschaft für Schwerionenforschung mbH (GSI) at Darmstadt, Germany}

%%%%%%%%%%%%%%%%%%%%%%%%%%%%%%%%%%%%%%%%%%%%%%%%%%%%%%%%%%%%%%%%%%%%%%%%%%%%%%%
\begin{abstract}
    The recent discovery of binary neutron star mergers has opened a new and exciting venue of research into hot and dense strongly interacting matter. For the first time this elusive state of matter, described by the theory of quantum chromo dynamics, can be studied in two very different environments. On the macroscopic scale in the collisions of neutron stars and on the microscopic scale in collisions of heavy ions at particle collider facilities. We will discuss the conditions that are created in these mergers and the corresponding high energy nuclear collisions. This includes the properties of QCD matter, i.e. the expected equation of state as well as expected chemical and thermodynamic properties of this exotic matter. 
    To explore this matter in the laboratory - a new research prospect is available 
    at the Facility for Antiproton and Ion Research, FAIR. The new facility is being constructed adjacent to the existing accelerator complex of the GSI Helmholtz Center for Heavy Ion Research at Darmstadt/Germany, expanding the research goals and technical possibilities substantially.
    The worldwide unique accelerator and experimental facilities
    of FAIR will open the way for a broad spectrum of unprecedented research supplying a variety of experiments in hadron, nuclear, atomic and plasma physics as well as biomedical and material science which will be briefly described.
\end{abstract}
%%%%%%%%%%%%%%%%%%%%%%%%%%%%%%%%%%%%%%%%%%%%%%%%%%%%%%%%%%%%%%%%%%%%%%%%%%%%%%%

\maketitle

\section{Introduction}
Until the recent discovery of the first binary neutron star merger \citep{TheLIGOScientific:2017qsa}, relativistic nuclear collisions were the only direct experimental access to study the properties of hot and dense strongly interacting matter. Our current theoretical knowledge of the thermodynamic properties of hot Quantum Chromo Dynamics (QCD) is very limited. Lattice QCD calculations consistently show a chiral crossover at vanishing chemical potential and at a pseudo-critical temperature of $\approx$ 155 MeV \citep{Aoki:2006we,Borsanyi:2013bia,Bazavov:2014pvz}. The interpretation of the chiral crossover and its exact relation to the deconfinement phenomenon is still not well understood. Since such direct calculations are restricted to vanishing baryo chemical potentials due to the infamous sign problem, not much is known on how this crossover continues into the high density region of the QCD phase diagram. In fact it is known that singularities in the imaginary plane of the phase diagram prevent us from expanding QCD thermodynamics to chemical potentials larger than $\pi > \mu_B/T$ \citep{Vovchenko:2017gkg,deForcrand:2002hgr,DElia:2002tig}. 
In the high density and very low temperature region of the phase diagram there is some guidance from cold compact stars: measurements of the mass-radius relation of compact stars could uniquely determine the equation of state~(EoS). However, such measurements are still scarce and have large uncertainties. 
The current method of exploring the properties and phase structure of high density and high temperature QCD matter
is to combine effective models of QCD with dynamical simulations of relativistic nuclear collisions. It is then attempted to relate the measurable hadron multiplicities, the fluctuations thereof and correlations between these final state hadrons to features in the phase diagram.

\begin{figure}[t]
%    \begin{center}
   \includegraphics[width=0.5\textwidth]{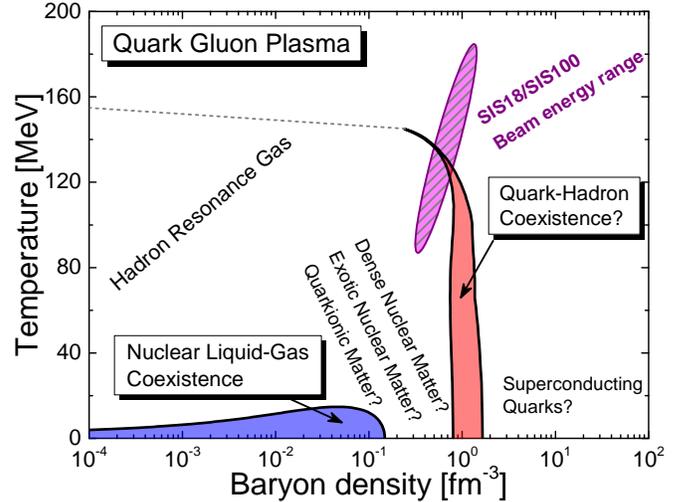}
    \caption{Conjectures phase diagram of QCD. The nuclear liquid gas transition is shown as blue shaded area and a possible quark-hadron coexistence as red shaded area. The current and future SIS accelerators cover the only beam energy range where this transition can be studied in experiment.}
    \label{pd_draw}
%\end{center}
\end{figure}

An example of such a conjectured phase diagram is shown in figure~\ref{pd_draw}. Here the different possible phases of QCD and the transitions between these phases are shown as functions of the temperature and baryon density.

Over the last decades many such observables have been proposed. For example, it is conjectured that the so-called directed flow, which quantifies the dipole moment of the transverse momentum distribution as a function of the rapidity in nuclear collisions, should be sensitive to the initial compression of such a collision and thus to the equation of state \citep{Stoecker:1980vf,Brachmann:1999xt}. Other observables have been related to higher moments of the transverse flow. 
On the other hand it is known that phase transitions can induce clustering \citep{Chomaz:2003dz,Randrup:2003mu,Sasaki:2007db,Steinheimer:2012gc} which could lead to observable correlations in phase space. The existence of a critical endpoint may lead to a strong enhancement of the correlation length and therefore also increased particle number fluctuations. Finally, first attempts have been made to employ methods of machine- and deep-learning to create new features which may be sensitive to the properties of the QCD equation of state \citep{Pang:2016vdc,Steinheimer:2019iso,Du:2019civ}.

All above methods have yet to conjure any robust proof for the existence of a chiral or deconfinement phase transition. This ambiguity is often due to the uncertainties in the dynamic description of nuclear collisions. Since the system is small and very short-lived, effects of non-equilibrium and incomplete thermalization can be very important. Yet, the most convenient way to study the QCD EoS is to use relativistic fluid dynamic simulations that incorporate an effective EoS in addition to extensions that allow for the description of phase separation of critical dynamics.

The results of these simulations are then contrasted with experimental data from large international facilities like FAIR, which is currently under construction at GSI.

The new international facility FAIR (Facility for Antiproton and Ion Research in Europe \citep{bib1a,bib1b,bib1c,Sturm:2010yit,Stocker:2015cva})
is being constructed adjacent to the site of the GSI Helmholtz Center for Heavy Ion Research at Darmstadt, Germany. It will substantially expand research goals by providing unique accelerator
and detector facilities for a large variety of scientific applications. FAIR research focuses on the structure and evolution of cosmic matter on a microscopic scale. This will deepen our understanding of fundamental questions as:
\begin{itemize}
    \item How does the structure of matter arise from its basic constituents and the fundamental interactions?
    \item How can the structure of hadronic matter be deduced from the strong interaction? In particular, what is the origin of the hadron masses?
    \item What is the structure of matter under the extreme conditions of temperature and density also found in astrophysical objects?
    \item What was the evolution and the composition of matter in the early Universe?
    \item What is the origin of elements in the Universe?
\end{itemize}

To address these fundamental questions FAIR will provide an extensive range of particle beams from protons
and their antimatter partners, antiprotons to ion beams of all chemical elements up to the heaviest (stable)
one, uranium, with world record intensities. As a joint effort of more than 50 countries the new facility builds
and substantially expands, on the present accelerator system at the GSI Helmholtz Centre for Heavy Ion Research. Both,
in its research goals and its technical possibilities. Compared to the present GSI facility, the beam intensities will
increase by a factor of 100 for primary beams and up to a factor of 10000 for secondary radioactive beams with
an excellent beam brilliance of primary as well as secondary beams. This will be achieved through innovative
beam handling techniques, many aspects of which have been developed at GSI over recent years with the
present accelerator complex. This includes in particular stochastic, laser and electron-beam cooling of high-energy,
high-charge state ion beams in storage rings and bunch compression techniques. To realize a parallel operation of
the research programs the design of FAIR includes the superconducting double-ring synchrotrons SIS100 and SIS300.
Both have a circumference of 1100 meters each and a magnetic rigidity of 100 and 300 Tm, respectively. Since
a key feature of the new facility will be the generation of intense, high-quality secondary beams,
the facility design contains a system of associated storage rings for beam collection, cooling,
phase space optimization, and experimentation.

The {\it Modularized Start Version} \citep{bib1c,bib1d}
of FAIR consists of the superconducting synchrotron SIS100 which feeds primary beams for fixed-target
experiments and production targets for secondary beams of rare isotopes and antiprotons with
the two cooler-storage rings HESR (High Energy Storage Ring) and CR (Collector Ring). Detector areas
for all scientific pillars are included. Following an upgrade for high intensities, the existing GSI
accelerators UNILAC and SIS18 will serve as injectors. The facility will be completed by experimental
storage rings enhancing capabilities of secondary beams and by the superconducting synchrotron SIS300
providing parallel operation of experimental programs as well as particle energies twenty-fold higher
compared to those achieved so far at GSI.

\subsection{CBM and HADES -- A glimpse into the interior of neutron stars}
The stability of neutron stars is guaranteed by the Pauli principle, stabilizing the nucleons from gravitational collapse, together with the repulsive part of the strong nucleon-nucleon interaction. Therefore the structure of a neutron star is dictated by the strong interaction
with the nuclear equation-of-state as a key ingredient. At the transition from the outer to the inner core
of a neutron star exotic phases of matter can appear. This can include matter with additional and new hadron species carrying strangeness such as hyperons. Even pion or kaon condensed matter or deconfined quark matter is possible to exist here.
The question of whether the interior of a neutron star consists already of a (superconducting) quark phase is
an active field of research. To explore this kind of nuclear matter at super-saturation densities in the laboratory,
relativistic nucleus-nucleus reactions provide the only possibility.

The mission of high-energy nucleus-nucleus collision experiments worldwide is to investigate the properties
of strongly interacting matter under these extreme conditions. At very high collision energies, as available
at the RHIC and LHC, the measurements concentrate on the study of the properties of QCD matter
at very high temperatures and almost zero net baryon densities, trying to determine the features of deconfinement. The experimental confirmation of a possible phase transition at high net baryon density would be
a substantial progress in the understanding of the properties of strongly interacting matter.

Complementary to high-energy nucleus-nucleus collision experiments at the RHIC and LHC, the CBM experiment
\citep{bib3b,bib3c,bib3d} as well as HADES \citep{bib3i,bib3j,bib3k,bib3l,bib3m,bib3n} at SIS100/300
will explore the QCD phase diagram in the region of very high baryon densities and moderate temperatures. It will do so
by investigating heavy-ion collision in the beam energy range 2~-~35 AGeV \footnote{light nuclei up to 44 AGeV}.
This approach includes the study of the nuclear matter equation-of-state, the search for new forms of matter,
the search for the suggested first order phase transition to the deconfined phase at high baryon densities,
the QCD critical endpoint, and the chiral phase transition.
In the case of the first order phase transition, basically one has to search for non-monotonic
behavior of observables as function of the collision energy and system size. The CBM experiment at FAIR
is being designed to perform this search with a large range of observables, including very rare probes
at these energy regime like charmed hadrons. Produced near threshold, their measurement might be well suited
to discriminate hadronic from partonic production scenarios \citep{Steinheimer:2016jjk}.

The properties of hadrons are expected to be modified in a dense hadronic environment which is
eventually linked to the onset of chiral symmetry restoration at high baryon densities
and/or high temperatures. The experimental verification of this theoretical prediction is a challenging question. The dileptonic decays of light vector mesons ($\rho$,$\omega$,$\phi$) provide the tool to study such modifications since
the lepton daughters do not undergo strong interactions and can therefore leave the dense hadronic medium
essentially undisturbed by final-state interactions. For these investigations the $\rho$ meson plays
an important role since it has a short lifetime and a large probability to decay inside
the reaction zone when created in a nucleus-nucleus collision. As a detector system dedicated
to high-precision di-electron spectroscopy at beam energies of 1~--~2 AGeV, the modified HADES detector
at SIS100 will measure $e^+e^-$ decay channels as well as hadrons \citep{bib3l,bib3m,bib3n} in collisions
of light nuclei up to 10 AGeV beam energy. Complementary, the CBM experiment will cover
the complete FAIR energy range in collisions of heavy nuclei by measuring both the $e^+e^-$
and the $\mu^+\mu^-$ decay channels.

Most of the rare probes like lepton pairs, multi-strange hyperons and charm will be measured
for the first time in the FAIR energy range. The goal of the CBM experiment as well as HADES is to study
rare and bulk particles including their phase-space distributions, correlations and fluctuations
with unprecedented precision and statistics. These measurements will be performed in nucleus--nucleus,
proton--nucleus, and proton--proton collisions at various beam energies. The unprecedented beam intensities
will allow studying extremely rare probes with high precision which have not been accessible
by previous heavy-ion experiments at the AGS and the SPS.

\section{How to connect FAIR and neutron star mergers}

In the following, we will discuss how the experimental program of the CBM and HADES experiments at FAIR can be augmented by observation of gravitational waves from binary neutron star mergers. In particular we will discuss the properties of matter created in nuclear collisions and how it can be compared to properties of matter in such mergers. This will give some first indications on how these two very different experimental observations may be able to complement each other in the understanding of the equation of state of QCD.

\subsection{Constructing an effective EoS - the CMF model}

The binding element between the physics of heavy ion collisions and neutron star mergers is the QCD equation of state at large densities. To make any connections or even predictions in both scenarios one requires a consistent description of the properties of matter up to several times nuclear saturation density and Temperatures of the order of 100 MeV.

Since such conditions cannot be calculated with first principle methods, we will present an alternative approach. This approach is the Frankfurt Chiral Mean field Model (CMF) which is based on the fundamental symmetries of QCD \citep{Papazoglou:1998vr,Detar:1988kn}. The CMF model contains the appropriate degrees of freedom in the corresponding limits, i.e. confided hadronic matter \citep{Tanabashi:2018oca} and deconfined quarks and gluons. It also provides a good description of nuclear physics, lattice QCD results as well as cold neutron star properties.\\

The main component of the CMF model is a three flavor chiral Lagrangian \citep{Papazoglou:1998vr}. The Lagrangian contains:
\begin{eqnarray}
{\cal L}_{SU(3)_f}={\cal L_B}  + U_{\rm sc} + U_{\rm vec}
\end{eqnarray}
where ${\cal L_B}$ describes scalar and vector mean field interactions among the ground state octet baryons and their parity partners:
\begin{eqnarray}
{\cal L_B} &=& \sum_b (\bar{B_b} i {\partial\!\!\!/} B_b)
+ \sum_b  \left(\bar{B_b} m^*_b B_b \right) \nonumber \\ &+&
\sum_b  \left(\bar{B_b} \gamma_\mu (g_{\omega b} \omega^\mu +
g_{\rho b} \rho^\mu + g_{\phi b} \phi^\mu) B_b \right) \,,
\label{lagrangian2}
\end{eqnarray} 
The index $b$ runs over all ground-state baryons, $p$, $n$, $\Lambda$, $\Sigma^{+,0,-}$, $\Xi^{0,-}$, and their respective parity partners, $N(1535)^{+,0}$, $\Lambda(1405)$, $\Sigma(1750)^{+,0,-}$, $\Xi(1950)^{0,-}$.

$U_{\rm sc}$ describes the potential of the scalar $\sigma$ and $\zeta$ fields, and $U_{\rm vec}$ is the potential of the vector $\omega$, $\rho$, and $\phi$ fields.

The effective masses of the ground state  octet baryons and their parity partners read~\citep{Steinheimer:2011ea}:
\begin{eqnarray}
m^*_{b\pm} &=& \sqrt{ \left[ (g^{(1)}_{\sigma b} \sigma + g^{(1)}_{\zeta b}  \zeta )^2 + (m_0+n_s m_s)^2 \right]} \nonumber \\ 
& \pm & g^{(2)}_{\sigma b} \sigma \ ,
\end{eqnarray}
where the coupling constants $g^{(*)}_{*b}$ are determined by the vacuum masses and by nuclear matter properties. $m_0$ refers to a bare mass term of the baryons which is not generated by the breaking of chiral symmetry, and $n_s m_s$ is the SU(3) breaking mass term that generates an explicit mass corresponding to the strangeness $n_s$ of the baryon. 

The vector interactions lead to a modification of the effective chemical potentials for the baryons and their parity partners:

\begin{equation}
    \mu^*_b=\mu_b-g_{\omega b} \omega-g_{\phi b} \phi-g_{\rho b} \rho
\end{equation}

This approach describes chiral restoration through parity doubling in the baryon octet \citep{Detar:1988kn,Zschiesche:2006zj,Aarts:2017rrl,Sasaki:2017glk}. As a consequence the effective nucleon mass never drops significantly below its vacuum expectation value.

The chiral field dynamics are determined self consistently by the scalar meson interaction potentials.
The mean field vector repulsion is mediated by the fields: $\omega$ for repulsion at finite baryon densities, the $\rho$ for repulsion at finite isospin densities, and the $\phi$ for repulsion when finite strangeness density is present. 

The coupling strengths of the nucleons and hyperons were chosen to reproduce nuclear binding energies as well as optical potentials in nuclear matter.

\begin{figure*}[t]
    \begin{center}
        \includegraphics[width=1.0\textwidth]{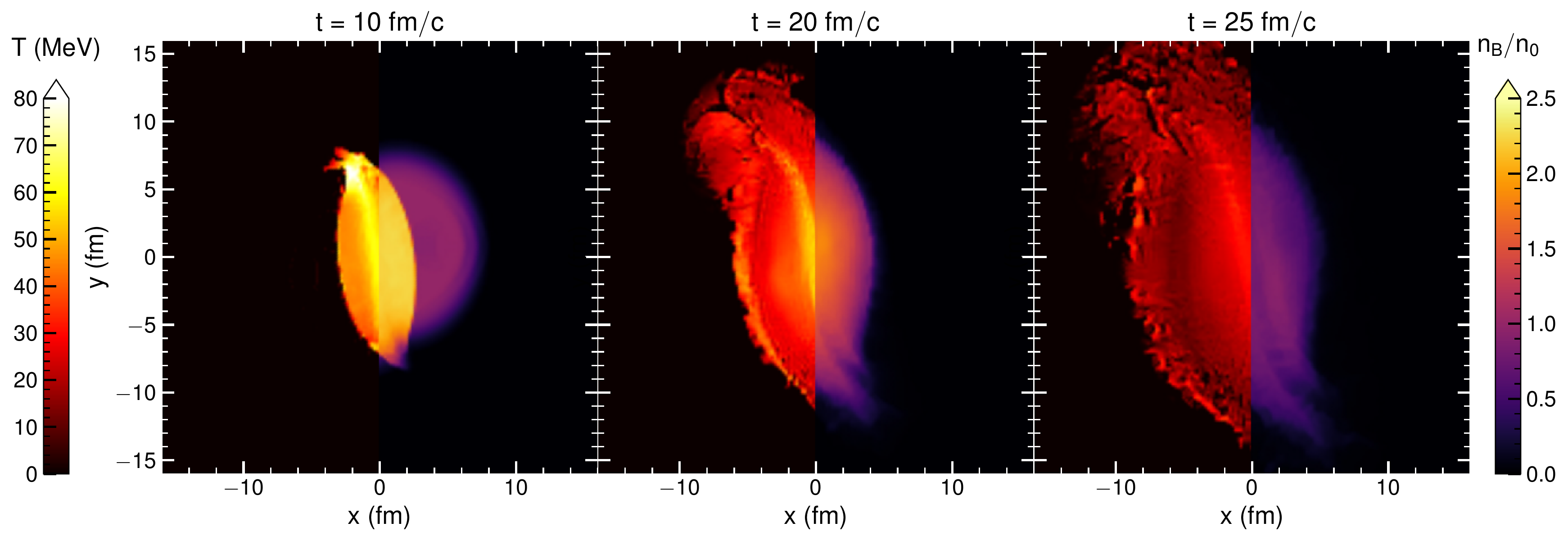}
    \caption{Temperature (left part) and density (right part) distributions at different times for a Pb+Pb collision at $\mathrm{E}_{\mathrm{lab}}=0.6 \ A$ GeV. The simulations were done using a fully relativistic ideal fluid dynamic solver. The largest temperatures are achieved directly at the shock front where the two incoming nuclei make the first contact. The whole collision duration is only about 25 fm/c after which the system decouples and hadrons freeze out. }
    \label{3dtime}
\end{center}
\end{figure*}

The remaining mesonic and non-interacting hadronic degrees of freedom are included in a form of Hadron Resonance Gas as a thermal heat bath according to their vacuum masses.

The baryonic interactions allow for a reasonable description of nuclear matter properties.

The quark degrees of freedom are introduced as in the Polyakov-loop-extended Nambu Jona-Lasinio (PNJL) model~\citep{Fukushima:2003fw,Motornenko:2019arp}:
\begin{eqnarray}
	\Omega_{\rm q}=&-&VT \sum_{q_{i}\in Q}\frac{d_{q_{i}}}{(2 \pi)^3}\int{d^3k} \frac{1}{N_c}\ln\left(1+3\Phi e^{-\left(E_{q_{i}}^*-\mu^*_{q_{i}}\right)/T}\right.\nonumber\\
	&+&\left.3\bar{\Phi}e^{-2\left(E_{q_{i}}^*-\mu^*_{q_{i}}\right)/T} +e^{-3\left(E_{q_{i}}^*-\mu^*_{q_{i}}\right)/T}\right)\,,
	\label{eq:q}
\end{eqnarray}
where the index ${q_{i}}$ runs through $u,d,s$ flavors. The anti-quark contribution can be obtained by replacing $\mu^*_{q_{i}}\rightarrow-\mu^*_{q_{i}}$, and $\Phi\leftrightarrow\bar{\Phi}$. The Polyakov-loop order parameter $\Phi$ effectively describes the gluon contribution to the thermodynamic potential and is controlled by the temperature dependent potential~\citep{Motornenko:2019arp}:

\begin{eqnarray}
    U_{\rm Pol}(\Phi,\overline{\Phi},T) &=& -\frac12 a(T)\Phi\overline{\Phi} \\ \nonumber
	 + b(T)\ln \Bigl[1-6\Phi\overline{\Phi}\Bigr.
	 &+& \Bigl. 4(\Phi^3+\overline{\Phi}^{3})-3(\Phi\overline{\Phi})^2\Bigr] \,, \\
    a(T) &=& a_0 T^4+a_1 T_0 T^3+a_2 T_0^2 T^2,  \nonumber \\
    b(T) &=& b_3 T_0^4  \nonumber\,.
\end{eqnarray}

The dynamical quark masses $m_q^*$ of the light and strange quarks are also determined by the $\sigma$- and $\zeta$- fields, with the exception of a fixed mass term $m_{0q}$:
\begin{eqnarray}
m_{u,d}^* & =-g_{u,d\sigma}\sigma+\delta m_{u,d} + m_{0u,d}\,,&\nonumber\\
m_{s}^* & =-g_{s\zeta}\zeta+ \delta m_s + m_{0q}\,.&
\end{eqnarray}

The full grand canonical potential of the CMF model is then:
\begin{eqnarray}
\Omega=\Omega_{\rm q} + \Omega_{\bar{\rm q}} + \Omega_{\rm h} + \Omega_{\rm \bar{h}}  - \left(U_{\rm sc} + U_{\rm vec} + U_{\rm Pol}\right)\,,
\end{eqnarray}
$\Omega_{\rm h}$ and $\Omega_{\rm \bar{h}}$ are the contributions from the hadrons. $U_{\rm sc}$ is the mean field interaction potential of the scalar mean fields $\sigma$ and $\zeta$, and  $U_{\rm vec}$ of the repulsive vector mean fields $\omega$, $\rho$, and $\phi$. $U_{\rm Pol}$ describes an effective gluon potential contribution as a part of the PNJL description.

The transition between the quark and hadronic degrees of freedom is controlled by two mechanisms.
As the Polyakov loop order parameter becomes finite, free quarks  appear.
The suppression of hadrons at high energy densities is maintained by their excluded-volume hard core interactions~\citep{Rischke:1991ke,Steinheimer:2011ea}.

The CMF model predicts two first order phase transitions for isospin symmetric matter. The nuclear liquid-vapor phase transition has a critical temperature $T_{\rm CP}\approx 17$ MeV.
At higher densities, the CMF model has another first order phase transition due to the chiral symmetry restoration~\citep{Steinheimer:2011ea,Motohiro:2015taa} with rather low critical temperature $T_{\rm CP}\approx 17$ MeV. 

The CMF model can readily be applied to study neutron stars without changing any of its parameters. Only electric charge neutrality and $\beta$-equilibrium are imposed so the conditions of the neutron star interior are fulfilled.

A more detailed description of the CMF model with its parameters can be found in~\citep{Motornenko:2020yme}.

\section{Fluid dynamic description of nuclear collisions at GSI and FAIR}

The most convenient and consistent way to study the QCD equation of state in heavy ion collisions (as well as neutron star mergers) is through the use of relativistic fluid dynamics. In this approach the equation of state enters naturally as a necessity to close the system of the fluid dynamic equations. For the current study we describe the time evolution of the system based on the equations of relativistic ideal fluid dynamics, namely local four-momentum conservation,
\begin{equation}
\partial_{\mu} T^{\mu \nu} =0\ ,
\end{equation}
and local flavor conservation,
\begin{equation}\label{consr}
\partial_{\mu} j^{\mu}=0\ .
\end{equation}
We include only the net baryon number current.
These equations can be solved numerically on a 3+1 dimensional Cartesian lattice \citep{Rischke:1995ir}.

For the SIS beam energy range, different methods of constructing the initial state of the fluid simulations have been proposed \citep{Petersen:2008dd,Karpenko:2013wva,Akamatsu:2018olk}. It has been shown that for such low beam energies the initial compression is very dependent on the equation of state \citep{Rischke:1995mt}. To include this important compression phase in the fluid dynamic evolution, we initialize the system as two Lorenz contracted Wood-Saxon distributions with a small offset along the transverse-axis, i.e. the collision has an impact parameter of $b=2$ fm. The time evolution of the temperature and net baryon density in the event plane, i.e. the plane between the beam axis (x-axis in the figure) and impact parameter direction (y-axis in the figure) is shown in figure \ref{3dtime}. Here the temperature (left) and density (right) are shown for three different times. One important observation is that the density and Temperature distribution is rather homogeneous in the collision zone, except for a small region where the two incoming nuclei first overlap and form a shockwave.

We find that the whole system formed has a very narrow spread in the entropy per baryon both of which are conserved. In addition the initial compression and excitation are very close to what is expected from a simple one dimensional solution of the shock-wave problem, called the Rankine-Hugoniot-Taub adiabat (RHTA) solution:

\begin{equation}
    \label{eq:Taub}
    (P_0+\varepsilon_0)\, (P+\varepsilon_0)\, n^2=(P_0+\varepsilon)\, (P+\varepsilon)\, n^2_0\,,
\end{equation}
where $P_0$, $\varepsilon_0$ and $n_0$ correspond to the initial pressure, energy density, and baryon density in the local rest frame of each of the two colliding slabs \citep{Stoecker:1978jr,Stoecker:1980uk, Stoecker:1981za, Stoecker:1981iw,Hahn:1986mb,Stoecker:1986ci,Merdeev:2011bz}. The two incoming nuclei consist of nuclear matter in the ground state, $P_0=0,~\varepsilon_0/n_0 - m_N=-16$ MeV and $n_0=0.16~ {\rm fm^{-3}}$.
Furthermore, the collision energy can then be related to the created density as:
\begin{equation}
    \label{eq:stopping}
    \gamma^{\rm CM}=\frac{\varepsilon n_0}{\varepsilon_0 n},~\gamma^{\rm CM}=\sqrt{\frac{1}{2}\left(1+\frac{E_{\rm lab}}{m_N}\right)}\,.
\end{equation}
Here $\gamma^{\rm CM}$ is the Lorentz gamma factor in the center of mass frame of the heavy ion collisions and $E_{\rm lab}$ is the beam energy per nucleon in the laboratory frame of a fixed target collision.

Using this simple yet accurate solution we can estimate the initial densities as well as the totally produced entropy per baryon in heavy ion collisions at the SIS energy range.

\begin{figure}[t]
    \begin{center}
           \includegraphics[width=0.5\textwidth]{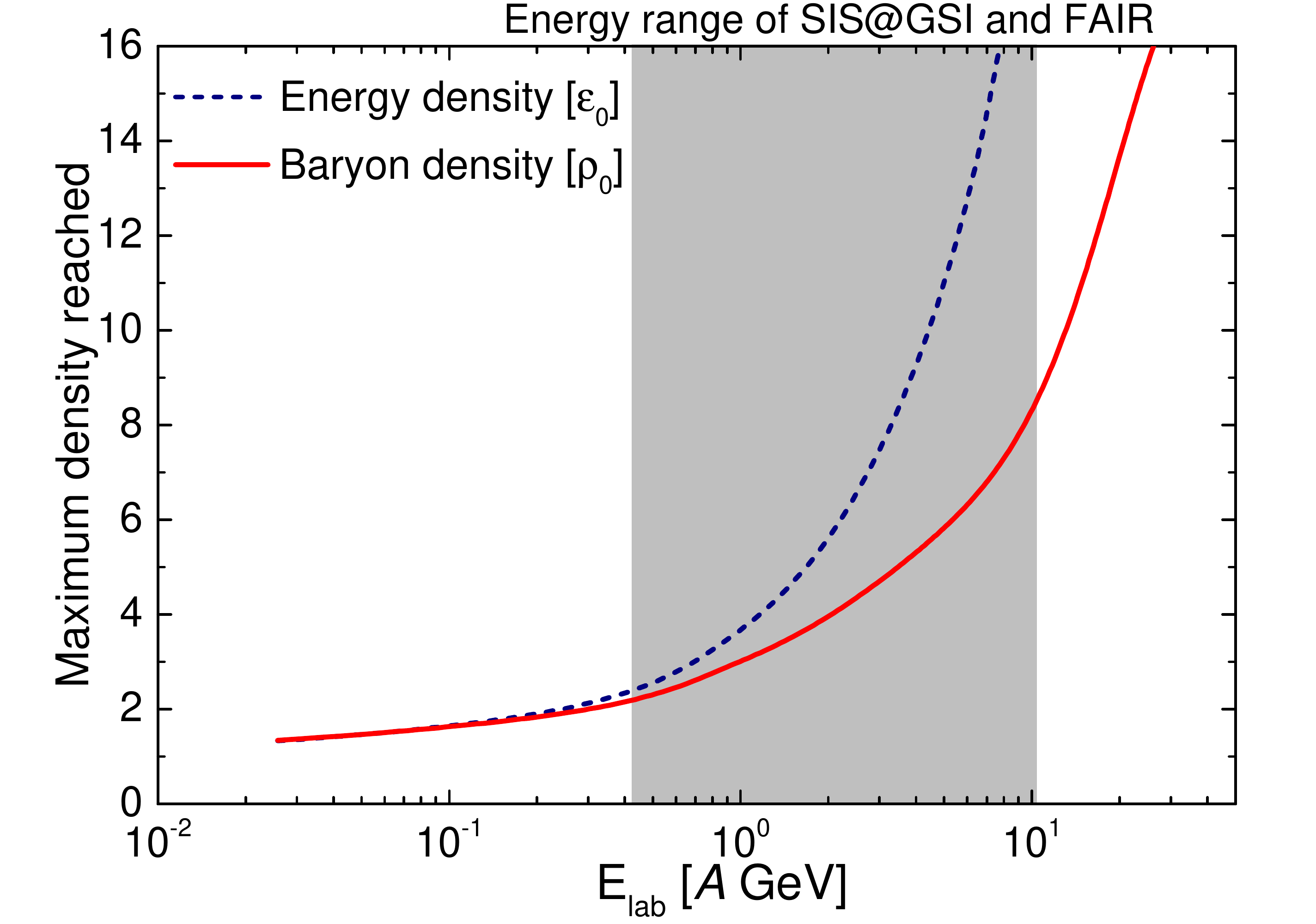}
    \caption{ Beam energy dependence of the highest energy (dashed line) and net baryon number (solid line) densities reached in central nuclear collisions. The densities are calculated self-consistently using the CMF model and the RHT-Adiabat solution. The beam energy range if the SIS accelerator is shown as grey band. 
    }
    \label{initial_rho}
\end{center}
\end{figure}

The expected initial energy and baryon number densities for central heavy ion collisions calculated this way, using the CMF model consistently, are shown in figure \ref{initial_rho}. The densities are scaled by their respective nuclear saturation values. It is clear that the SIS accelerator will cover a wide range of maximal compression between twice and almost twenty times nuclear saturation energy density and up to ten times nuclear saturation density. In this density range a lot of interesting phenomena, like the chiral transition and the deconfinement transition are expected to occur.

\begin{figure}[t]
    \begin{center}
           \includegraphics[width=0.5\textwidth]{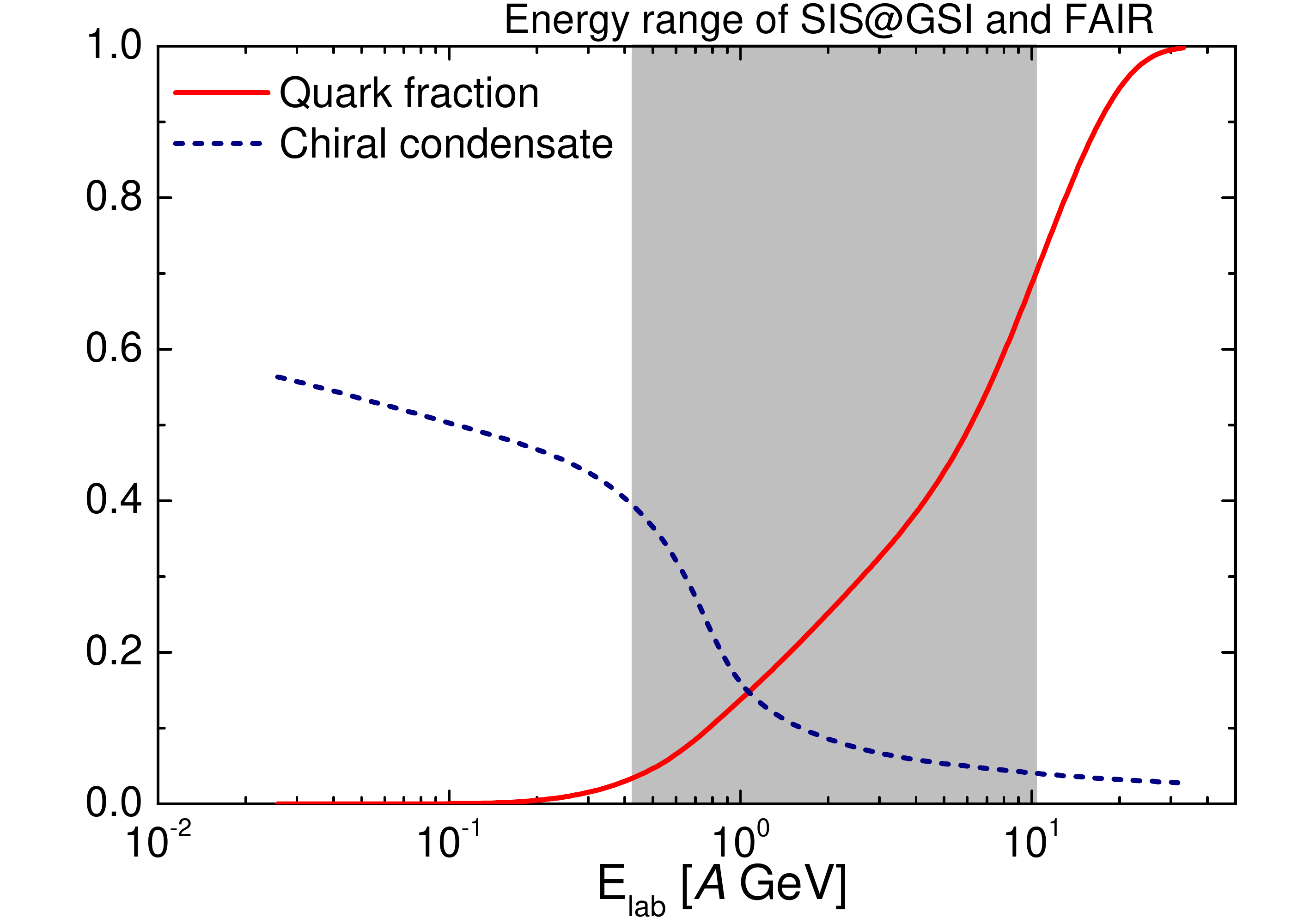}
    \caption{ Beam energy dependence of the chiral condensate (dashed line) and quark fraction of the baryon density (solid line) in central nuclear collisions. The values are calculated self-consistently using the CMF model and the RHT-Adiabat solution. The beam energy range if the SIS accelerator is shown as a grey band. 
    }
    \label{initial_chiral}
\end{center}
\end{figure}

To highlight this transition region, figure \ref{initial_chiral} shows the beam energy dependence of the chiral condensate $\sigma/\sigma_0$ from the CMF model as well as the fraction of the total baryon density which is made up by free quarks. Several interesting features can be observed.
\begin{itemize}
    \item The chiral condensate already for cold nuclear matter has dropped to a value significantly below unity which is a well known effect of the density dependence of the chiral condensate. 
    \item In the beam energy range of a few GeV per baryon, the chiral transition occurs and chiral symmetry is almost completely restored.
    \item The deconfinement of quarks occurs at slightly higher beam energies and is completed only for beam energies of the order of $\mathrm{E_{lab}}\approx 20 \ A$ GeV
\end{itemize}

\begin{figure}[t]
    \begin{center}
           \includegraphics[width=0.5\textwidth]{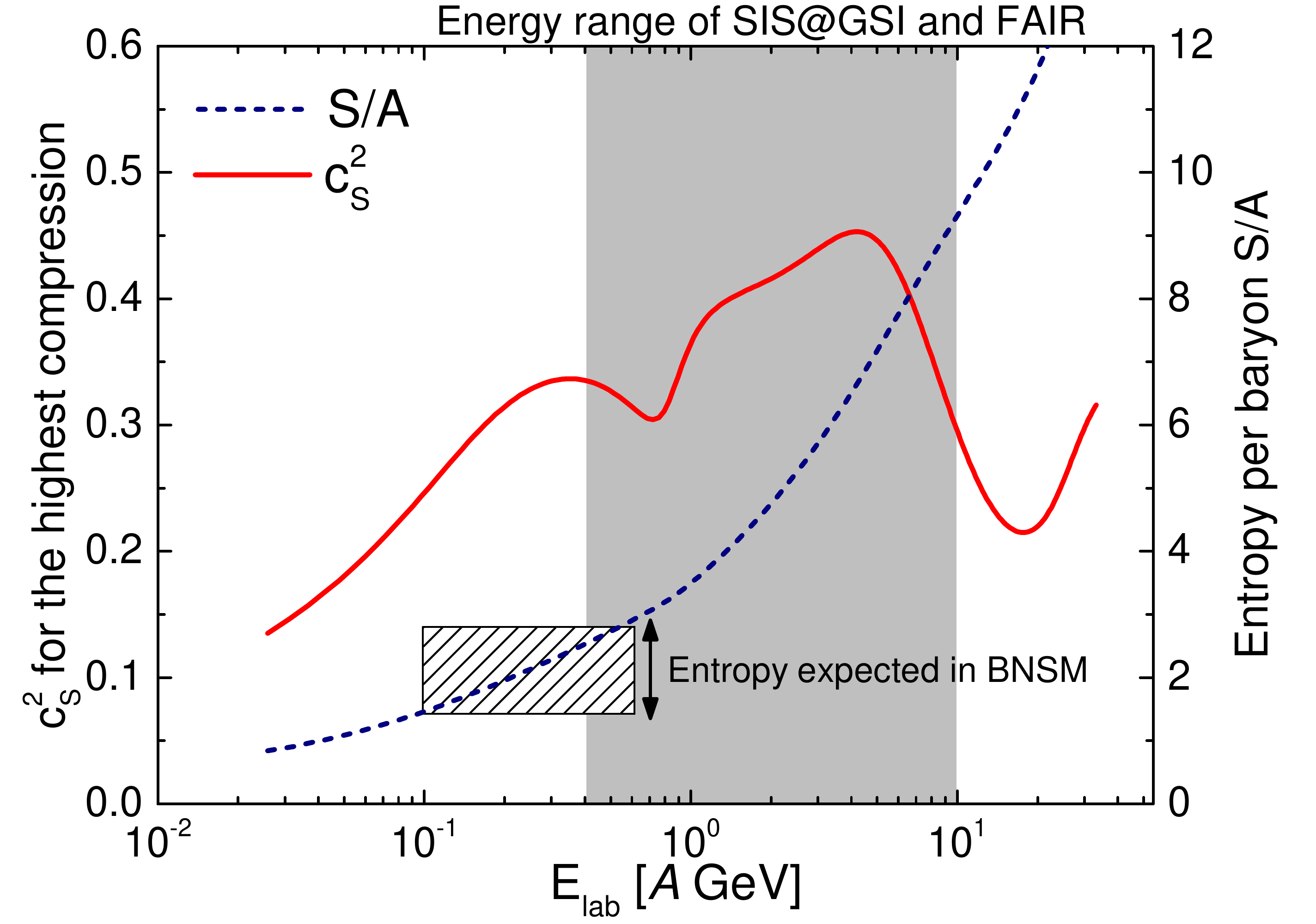}
    \caption{Beam energy dependence of the speed of sound (solid line) and specific entropy (dashed line) in central nuclear collisions. The values are calculated self-consistently using the CMF model and the RHT-Adiabat solution. The beam energy range if the SIS accelerator is shown as grey band. The expected range of specific entropies that can be reached in BNSM is indicated as small box.
    }
    \label{initial_cs2}
\end{center}
\end{figure}

These features make the SIS energy range extremely interesting to study. In addition the phase change will be reflected in the change of the transport properties of matter, in particular the speed of sound. Thus, we show the speed of sound for the highest compression as function of beam energy as red solid line in figure \ref{initial_cs2}. Both transitions, the chiral as well as deconfinement crossover are visible as minima in the speed of sound. Notably, the overall speed of sound is rather large in the whole SIS beam energy range. This is due to the fact that the matter here is still mainly composed of dense hadronic matter, i.e. baryons, which exhibit a strong short range repulsive interaction. This makes the equation of state very stiff, until the appearance of quarks and gluons soften the equation of state. The small softening due to the chiral transition is therefore not expected to lead to significant effects.

Finally, we also display the produced specific entropy (or entropy per baryon $S/A$) as function of beam energy as dashed line. The specific entropy per baryon is an important quantity since one usually estimates the consecutive expansion after the initial compression to be nearly isentropic. The final entropy per baryon is determined at the initial state and allows for a direct comparison with final state hadron multiplicities. The expected specific entropy for the SIS range is between 2 and 10. 

This determination of $S/A$ allows us to now also make a direct comparison of matter created at the SIS with matter created in a very different cosmic laboratory: the binary neutron star merger (BNSM). While a rough estimate of the expected maximum specific entropy in a BNSM is already indicated in figure \ref{initial_cs2} more detailed simulations are presented in \citep{Most:2019onn,ourpaper}. 
In the following we will simply assume that the specific entropy reached in BNSM can reach roughly 2.

\section{Properties of the hottest matter in neutron star mergers}
As shown in previous simulations of neutron star mergers, the hottest matter in these collisions is formed in a rotating ring around the dense and cold core of the newly formed massive star. Since the cold and dense matter can essentially be described by the same equations of state that also describe the well known static neutron stars, we will focus on the properties of the hot 'ring of fire' that surrounds the cold core. As mentioned before we can assume that the specific entropy in this ring, $S/A \approx 2$ \citep{ourpaper}. And since its evolution is described by ideal fluid dynamics the matter will evolve around an isentrope of that value. 

Figure \ref{bnsm_ch} shows the temperature dependence of several interesting quantities in the temperature range expected for this 'ring of fire', for $S/A=2$:

 \begin{figure}[t]
    \begin{center}
           \includegraphics[width=0.5\textwidth]{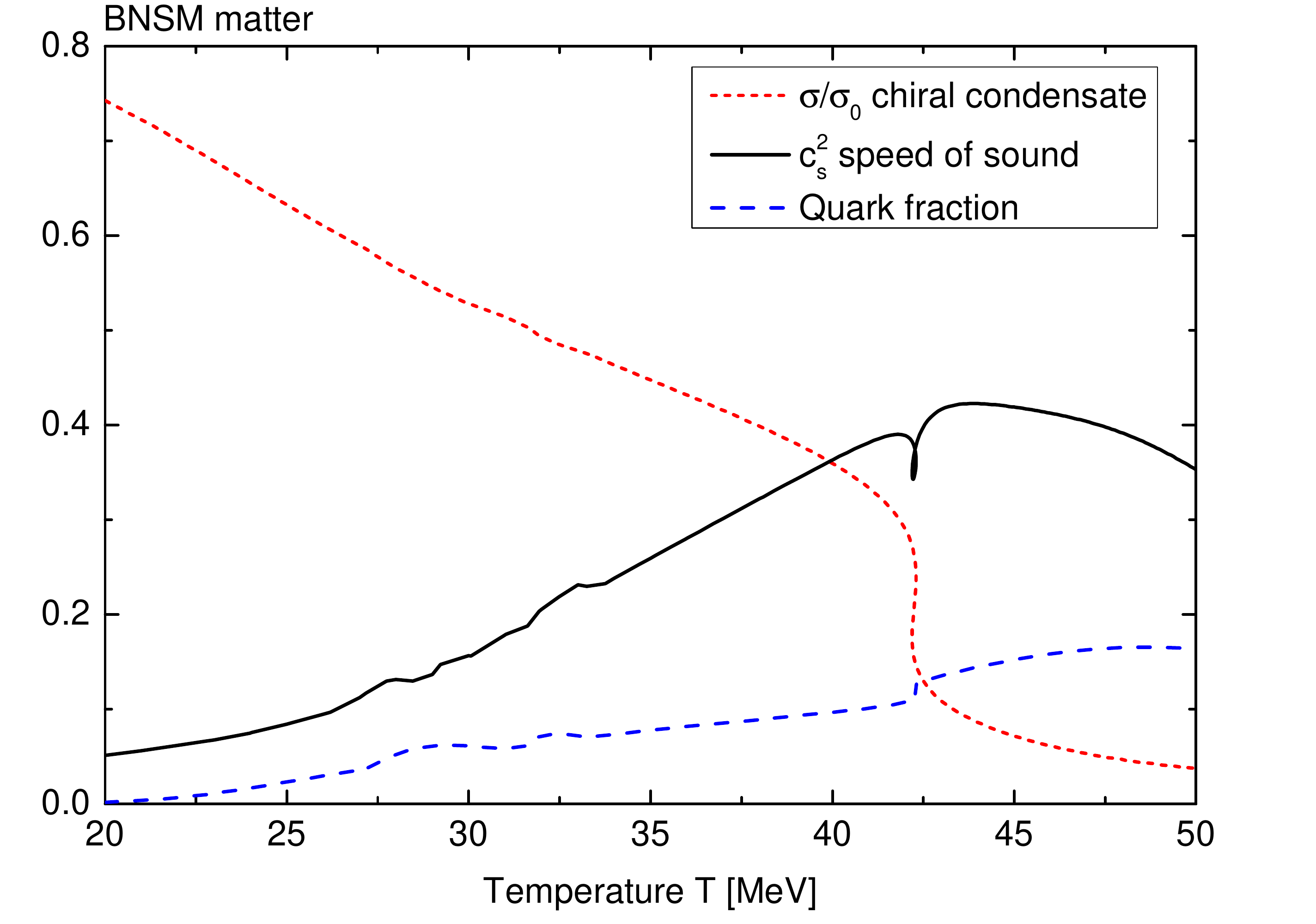}
    \caption{Temperature dependence of the chiral condensate (red dashed line) the speed of sound squared (black solid line) and the quark fraction (blue dashed line) for fixed specific entropy of $S/A=2$. The chiral crossover transition is reached at a temperature of $T\approx 42$ MeV.
    }
    \label{bnsm_ch}
\end{center}
\end{figure}

\begin{itemize}
    \item The normalized chiral condense $\sigma/\sigma_0$ (red dashed line). The value of the chiral condensate drops continuously due to the increase of the nucleon density as the temperature increases. Then at $T \approx 42$ MeV the chiral condensate exhibits a steep drop due to the chiral crossover occurring at that temperature. Since the study of the chiral transition is of great interest for the understanding of the QCD phase structure it is essential to understand whether this transition is actually reached in dynamical scenarios. The actual maximal temperature reached in this merger is therefore important.
    \item The speed of sound (black solid line). As in heavy ion collisions the speed of sound of QCD matter provides information on the equation of state. Similarly to heavy ion collisions, with increasing temperate the speed of sound increases due to the strong repulsion between the nucleons. This increasing trend only stops once the system reaches the chiral transition at which a small dip in the speed of sound is observed. Only at even higher temperatures when quark degrees of freedom become relevant the speed of sound decreases eventually towards the limit of an ideal gas. 
    \item The quark fraction of the QCD matter (blue dashed line). The quark fraction is defined as the fraction of the quark density over the total baryon density. In this scenario the quark fraction reaches at most $10 \%$ even for the highest temperatures. The hottest areas of the BNSM are therefore not composed of deconfined matter but of a very dense nuclear liquid close to chiral symmetry restoration. 
\end{itemize}

 \begin{figure}[t]
    \begin{center}
           \includegraphics[width=0.5\textwidth]{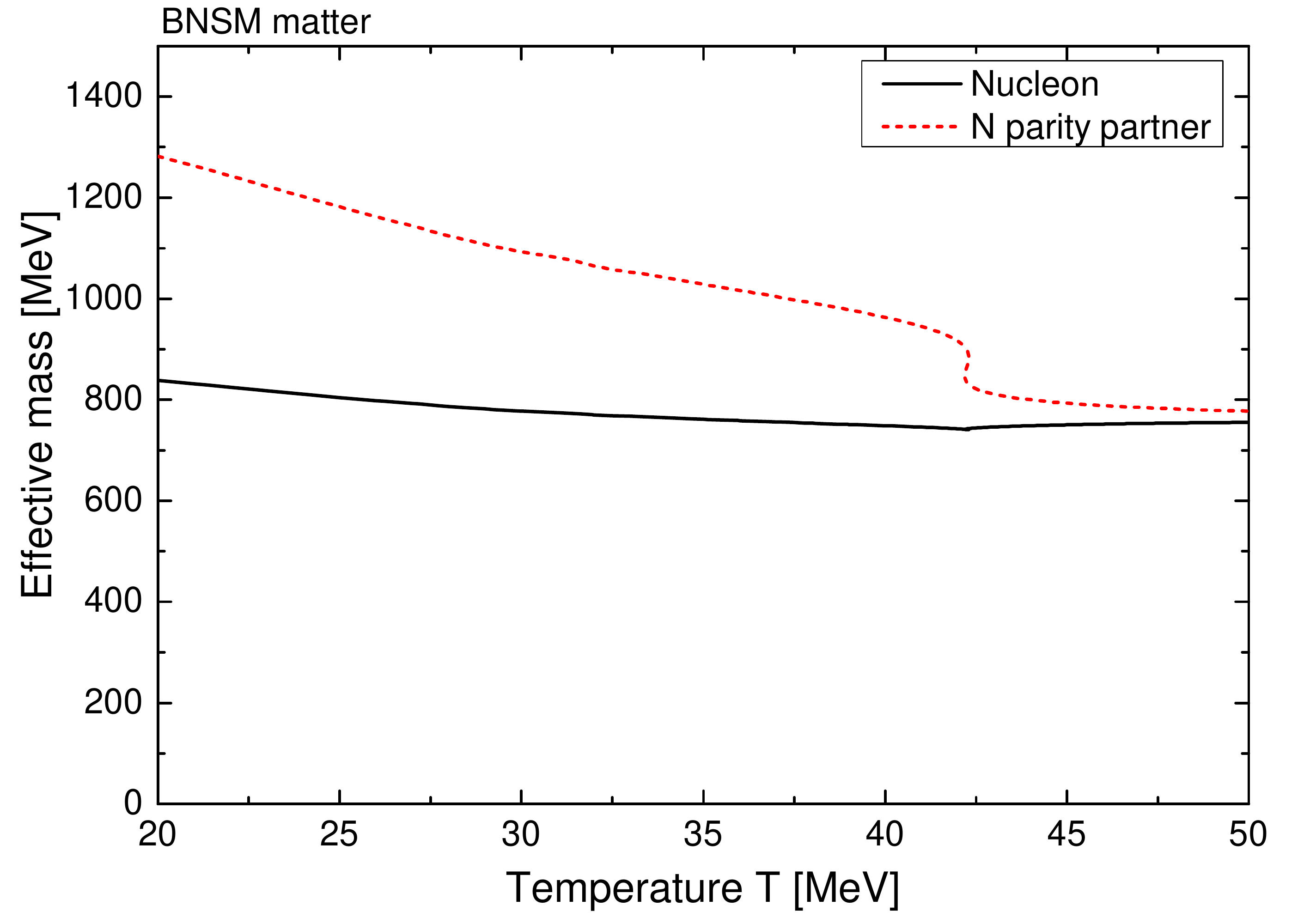}
    \caption{Temperature dependence of the effective masses of the nucleon (black solid line) and its parity partner (red dashed line) for a specific entropy of $S/A=2$. The chiral transition, where the two masses become approximately degenerate is observed at $T\approx 42$ MeV.
    }
    \label{bnsm_masses}
\end{center}
\end{figure}

In the parity doubling approach, the chiral transition manifests itself in the degeneracy of the ground state nucleon with its parity partner, once chiral symmetry is restored. This feature not only leads to an increase in the net baryon density along the transition but it also significantly alters the properties of the baryonic matter. This is of particular interest in heavy ion collisions where for example the emission of electromagnetic probes is sensitive to the medium properties \citep{Seck:2020qbx,Rapp:2009yu,Gale:2020xlg}. In neutron star mergers the electroweak processes may depend strongly on the modifications of hadrons in the dense and hot medium \citep{Dexheimer:2012eu}. The effect of, e.g., strongly varying effective mass of the nucleon parity partner will be an interesting topic for future studies. The dependence of the effective masses of the ground state nucleon and its parity partner on the temperature along the isentropic line is shown in figure \ref{bnsm_masses}. Here we observe that even though the two parity states become degenerate only at the chiral phase transition, the properties of the N$^*$ resonance will change drastically already at lower temperatures. Thus, it is interesting to consider possible effects of these in-medium properties on the evolution and emission during the BNSM.

\subsection{Important differences - strange matter}

In the previous sections we have shown how similar the matter in heavy ion collisions and the violent mergers of neutron stars can be and how both can be described consistently with the CMF model. 
Besides the important similarities there are also serious differences in these two scenarios. Since high energy nuclear collisions have lifetimes of the order of only tens of $\mathrm{fm}/c$, the production of new degrees of freedom, i.e. including strangeness, is often below its equilibrium value. In addition, strangeness can only be produced in $s$ + $\overline{s}$ pairs and strange quarks do not decay on these timescales. As a result strangeness is exactly conserved in nuclear collisions, leading to a finite strange chemical potential. In neutron star mergers, on the other hand, beta equilibrium breaks the conservation of strangeness and one usually assumes that $\mu_S=0$.

 \begin{figure}[t]
    \begin{center}
           \includegraphics[width=0.5\textwidth]{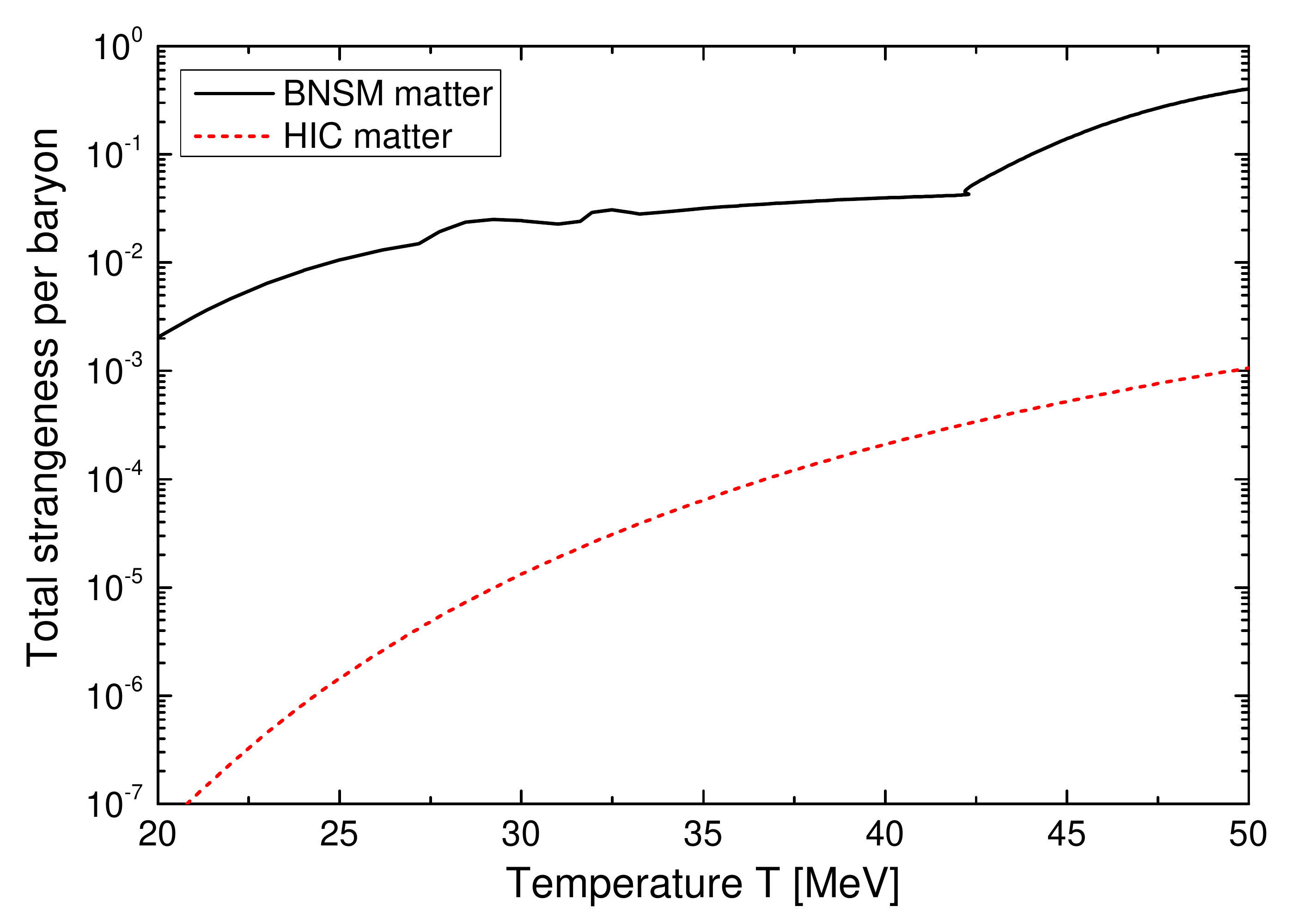}
    \caption{Temperature dependence of the total strangeness ($s + \overline{s}$) per baryon for isentropic trajectories $S/A \approx 2$ in nuclear collisions (red dashed line) and neutron star mergers (block solid line). It is obvious that the amount of strangeness produced in the two scenario differs by orders of magnitude. 
    }
    \label{strangeness_comp}
\end{center}
\end{figure}

The difference in the total strangeness (number of $s + \overline{s}$) per baryon in nuclear collisions and neutron star mergers is depicted in figure \ref{strangeness_comp}. Here, the total strangeness per baryon is shown for isentropic trajectories expected for BNSMs and nuclear collisions with $S/A=2$. The amount of strangeness in nuclear collisions is orders of magnitude smaller than in the neutron stars and can essentially be neglected during the evolution. In the BNSM the strangeness fraction increases and especially at the chiral transition starts to exceed ten percent of the total baryon number. The fact that deconfinement is not yet fully realized can also be observed since then the quark fraction would increase to unity. Overall we can conclude that strangeness in the hot 'ring of fire' of a BNSM plays a role though it does not dominate the dynamics. A fact that has to be considered when comparing  nuclear collisions at HADES and BNSMs.

\section{Summary}

It was shown that the matter created in binary neutron star mergers is very similar to the matter created in nuclear collisions at the SIS accelerator at the GSI and FAIR facility. This allows us to study the phase structure of QCD in two similar but also widely different setups. While nuclear collisions happen on a microscopic and neutron stars on a macroscopic scale, the equation of state which determines the ideal fluid dynamics used to describe both has the same features. This means that the equation of state of QCD is the binding piece between these two apparently different fields of physics. We have shown how one can construct an equation of state that satisfies the constraints of nuclear matter, neutron star matter, and even lattice QCD results at vanishing net baryon density in a single consistent approach. Using this CMF model we discussed similarities and differences between nuclear collisions and BNSMs which both can create matter with a specific entropy of $\approx 2$.

While the properties of the neutron star mergers will be investigated through gravitational waves measured by large interferrometers, the properties of nuclear collisions will be under investigation at the CBM and HADES experiments of the FAIR facility at GSI.

The construction of the Facility for Antiproton and Ion Research, FAIR, has started in 2013. The scientific program of FAIR covers
\begin{itemize}
\item The investigation of bulk matter in the dense and hot plasma state, atomic physics
of ultra-high electro-magnetic fields, material and biological research using intense,
highly-stripped ions and high-intensity laser fields (APPA).
\item The exploration of the QCD phase-diagram and the phase transition to
the deconfinement phase at high baryon densities in high-intensity nucleus-nucleus collisions (CBM and HADES).
\item The investigation of the nuclear chart and nuclear structure far from stability
and of nuclear astrophysics with radio-active beams (NuSTAR).
\item The exploration of the hadron structure, QCD vacuum, and the nature of the strong force
in the non-perturbative regime of quantum chromo dynamics with high-energy beams of antiprotons (PANDA).
\end{itemize}

Forming together the four scientific pillars. The design of FAIR includes the superconducting
double-ring synchrotrons SIS100 and SIS300 to realize parallel operation of the research programs.

Due to the high luminosity which exceeds current facilities
by orders of magnitude, experiments will be feasible that could not be done elsewhere. FAIR
will expand the knowledge in various scientific fields beyond current frontiers. Moreover, the
exploitation of exiting strong cross-topical synergies promises novel insights.
% ----------------------------------------------------------------------------------------------------------

\section*{Acknowledgments}
We thank E. Most, V. Dexheimer and L. Rezzolla for inspiring discussions about the features of BNSMs.
The authors are thankful for the support from HIC for FAIR and HGS-HIRe for FAIR.
JS thanks the Samson AG and the BMBF through the ErUM Data project for funding.
JS and HSt thank the Walter Greiner Gesellschaft zur F\"orderung der physikalischen Grundlagenforschung e.V. for its support.
HSt acknowledges the Judah M. Eisenberg Laureatus Chair at Goethe Universit\"at Frankfurt am Main.

\bibliography{Stoecker.bib}% Name of your .bib file

\end{document}